# A Polarization-insensitive and high-efficiency Schottky Photodetector integrated with a silicon ridge waveguide

Liu Yang[1], Pengfei Kou[1], Jianqi Shen[1], El Hang Lee[1], and Sailing He[1,2,*]

[1]Centre for Optical and Electromagnetic Research, Zhejiang Provincial Key Laboratory for Sensing Technologies, State Key Laboratory of Modern Optical Instrumentations, Zhejiang University, Hangzhou 310058, China

[2]ZJU-SCNU Joint Research Center of Photonics, South China Normal University, Guangzhou 510006, China

* Email: sailing@zju.edu.cn

**Abstract**－We propose a polarization-insensitive and high-efficiency plasmonic silicon Schottky diode for detection of sub-bandgap photons in the optical communication wavelength range through internal photoemission. Our photodiode is based on a hybrid plasmonic silicon waveguide. It has a gold film covering both the top and the sidewalls of a dielectric silicon waveguide with the Schottky contact formed at the gold-silicon interface. An extensive physical model is presented in detail and applied to calculate and analyze the performance of our detector. By comparison with a diode with only top contact of gold, the polarization sensitivity of responsivity is greatly minimized in our photodetector with sidewall coverage of gold. Much higher responsivities for both polarizations are also achieved in a very broad wavelength range of 1.2-1.5 μm. Moreover, the Schottky contact is only 4 μm long, leading to a very small dark current. Our design is very promising for practical applications in high-density silicon photonic integration.

Internal photoemission (IPE) is an intrinsic property of a Schottky diode, occurring at a metal-semiconductor interface [1]. In IPE, three processes are involved. Firstly, an electron (hole) in the metal is excited to a higher level after absorbing a photon, becoming a hot electron (hot hole). Secondly, the hot carrier (electron or hole) travels to the metal-semiconductor interface. During its travel, the hot carrier may lose some energy due to scattering by cold carriers or by thermal relaxation. Finally, upon arrival at the interface, it can cross over the Schottky barrier into the semiconductor if its remaining energy is still higher than the barrier. All the emitted hot carriers in the semiconductor will then be quickly swept away and collected by the Ohmic electrode as a photocurrent. In this case, only photons with energies larger than the Schottky barrier (even the energy is lower than the bandgap of the compositional semiconductor) can be converted into an electrical current. Therefore, IPE mechanism is always used to determine the Schottky barrier of a metal-semiconductor interface [1]. In recent years, IPE has been explored as one of the most attractive means to realize detection of photons below the semiconductor bandgap for sensing [2-4], solar energy conversion [5, 6], and optoelectronic devices for optical communication [7-11].

In particular, for planar silicon (Si)-based photodetectors, sub-bandgap photons in the optical communication wavelength range, i.e., 1.2-1.6 μm, can be easily detected by employing the IPE mechanism, since a metal (or silicide)-Si based Schottky diode usually has a much lower Schottky barrier than the Si bandgap energy (1.1 eV) [7-11]. This method is much simpler in comparison with previously reported methods, such as hybrid integration of other low-bandgap semiconductors [12], creating defect-induced low levels in the bandgap (which may cause high dark current) [13, 14], or employing the two-photon absorption effect (which requires high input power or high Q-factor cavities) [15]. However, the method has been known to have one shortfall. That is, the IPE probability of a Si-based Schottky diode, determined by the Schottky barrier, metal thickness, and the input photon energy, is not very high, leading to low internal quantum efficiency (IQE) and consequently the low responsivity [16]. In addition to choosing a metal (or silicide)-Si pair with a low Schottky barrier [7], increasing the metallic absorption as much as possible is another way to improve the responsivity. Plasmonic waveguides in the form of asymmetric [9] and symmetric [10] gold (Au) strips, or Au-capped photonic bus waveguide [11], have been proposed to form Schottky photodetectors, where the light waves propagate along the Au-Si interface and are absorbed more in the Au. However, these plasmonic photodetectors [9-11] are not suitable for practical applications in an optical interconnect system because of the intrinsic polarization-sensitive property of plasmonic strip waveguides [17]. Moreover, light waves are not well confined in these plasmonic waveguides, leading to long photodetectors (tens of micrometers) [9-11]. Therefore, the integration density is limited. The dark current, which is proportional to the metal-semiconductor contact area, is also very high.

In order to overcome these shortfalls, we propose in this paper, a new design. Our results show that the new diode is polarization-insensitive and is only 4-μm long, but still absorbs over 85% of light in the wavelength range of 1.2-1.6 μm. We have found that our photodetector shows greatly improved performance in comparison with the one reported in Ref. [11]. In this work, we extended the previously published physical model [16], which cannot be used directly, for the purpose of accurate calculation and analysis of the responsivity of the photodetectors presented here.

The schematic structure of our proposed photodetector is shown in Figure 1(a). It is integrated on a Si ridge waveguide so that the light propagates inside the waveguide can be detected directly by our photodetector. Our photodetector consists of a Si ridge waveguide, an Au Schottky electrode (of thickness $h_{Au}$) covering both the top and sidewalls of the Si ridge, an aluminium (Al) Ohmic electrode on top of the Si slab, and a thin SU8 insulating film (of thickness $h_{SU8}$) separating the two electrodes. The distinct feature, different from any of the previously reported structures [7-11], is the Au-covered Si ridge waveguide, which forms a special plasmonic waveguide. The Schottky contact appears at the Au-Si interface, as shown in Figure 1(b). The whole structure shown in Figure 1 is similar to the Si based hybrid plasmonic waveguide with low-index double silica nano-slots proposed in Ref. [19] for TE (transverse-electric) polarization. With the structure of Ref. [19], the silica nano-slots between Si and Au can be ultra-thin (e.g., < 2 nm) to form a metal-insulator-semiconductor Schottky diode. However, the optical field will be tightly confined in the silica nano-slots, which is not favorable for the mode absorption in Au or the coupling between the input Si ridge waveguide and the nano-slot hybrid plasmonic waveguide [19]. Without strong interaction between the input light and the Au film, the responsivity of the IPE-based Schottky diode is predicted to be limited. Alternatively, the nonlocal effect may appear, complicating the optical simulation [20], which is beyond the scope of this article. Thus, we will not consider the configuration of Ref. [19] but rather the one shown in Figures 1(a) and 1(b) with the Au film directly covering the Si waveguide surfaces. In this case, not only TE mode but also TM (transverse-magnetic) mode can be supported, and absorption for both polarizations can be achieved. The input Si ridge waveguide is designed as a single-mode dielectric waveguide with typical structural parameters of ridge width $w_r$ = 350 nm, ridge height $h_r$ = 250 nm, and slab thickness $h_{sl}$ = 50 nm. The top corners are rounded with a radius of $r$ to make the waveguide approach a real one (for example, fabricated by oxidation [11]). Since this parameter is not a main factor influencing the polarization sensitivity of our detector, we set it to a constant $r$ = 25 nm. Sub-bandgap photons guided in the Si ridge waveguide are directly coupled to the connected plasmonic one. Here, they are absorbed by the Au electrode and converted into photocurrent through the IPE effect. For comparison, we also considered the case when $h_{SU8}$ = 200 nm, as shown in Figure 1(c), which is similar to the Au-capped waveguide detector reported in Ref. [11]. The other structural parameters are kept the same as ours.

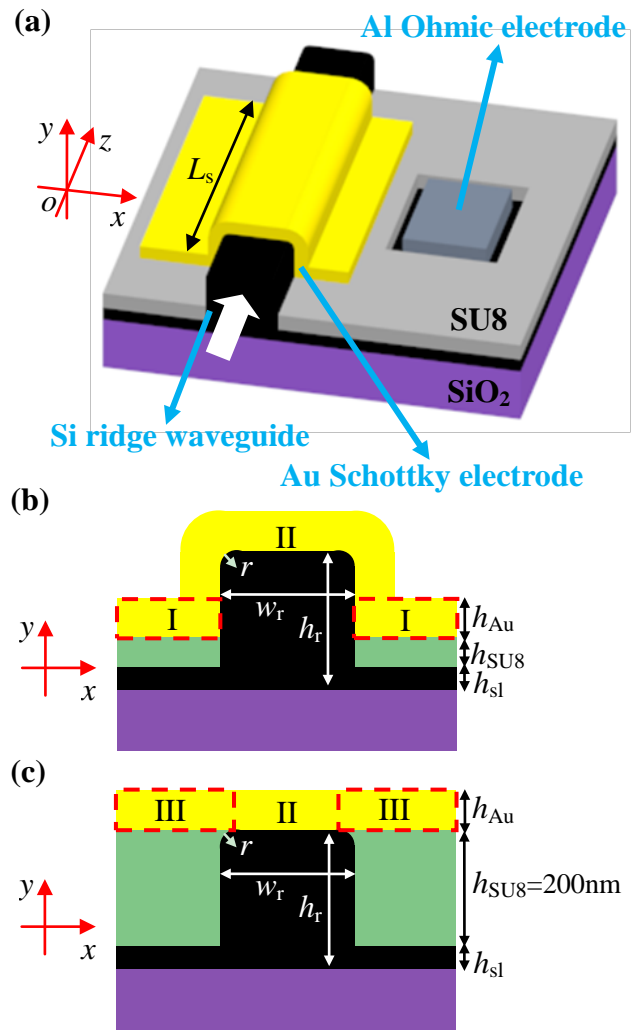

**Figure 1**. (a) Three-dimensional schematic diagram of our Si PD. Two dimensional schematic diagrams of our Si PD with (b) $h_{SU8}$ < 200 nm and (c) $h_{SU8}$ = 200 nm.

Generally, responsivity, $R$, for a PD in the short-circuit operation mode can be obtained directly from its internal quantum efficiency (IQE) as follows:

$$R = \frac{q}{E_0} \cdot A \cdot IQE \qquad (1)$$

where $A$ is the light absorption, $q$ is the elemental charge, and $E_0$ is the incident photon energy. For our IPE-based Schottky diode, IQE is dependent on the emission probability $P(E)$ of hot carriers, and can be obtained by integrating all emission probabilities in the energy range from the Schottky barrier $\Phi_B$ to the initial excess energy above the Fermi level $E_0$ [16] as follows:

$$IQE = \frac{1}{E_0} \int_{\Phi_B}^{E_0} P(E) dE \qquad (2)$$

$P(E)$ is a very important parameter for estimating IQE. Before deriving $P(E)$, we make two assumptions [16]: i) hot carriers which have no probability to emit through the Au-Si interface will be reflected elastically; and ii) those traveling to the Au-air or Au-SU8 interface cannot transmit through it but instead are totally and elastically reflected back. For a flat Au-Si interface, $P(E)$ has been derived in Ref. [16] and is expressed in Equations (3-1) and (3-2) according to the number of round trips the hot carriers can travel in the Au film.

$$P(E) = \begin{cases} P_i = \frac{1}{2}(1-\cos\Omega); & \text{when } N = 0; \quad (3\text{-}1) \\ P_{ii} = P_0 + (1-P_0)P_1 + (1-P_0)(1-P_1)P_2 + \cdots + P_N \prod_{m=0}^{N-1}(1-P_m); & \text{when } N > 0, \quad (3\text{-}2) \end{cases}$$

where $\Omega$ is the solid angle determined by $k \cdot \cos\Omega = k_{SB}$, leading to $\cos\Omega = \sqrt{\Phi_B/E}$ as shown in Figure 2(a); $P_{x\,(0-N)} = P(E_x) = \frac{1}{2}\left(1-\sqrt{\frac{\Phi_B}{E_x}}\right)$ is the emission probability of a hot carrier of energy $E_x$ which is reduced from $E_0$ to $E_0 e^{-2xh_{Au}/L}$ after travelling an $x$-number of round trips within the Au film. $N = \frac{L}{2h_{Au}}\ln\left(\frac{E_0}{\Phi_B}\right)$ is an integer, representing the number of round trips a hot carrier can travel back and forth in the Au film before its energy $E_0$ is reduced to $\Phi_B$ and has no chance to jump over the Schottky barrier; $L$ is the mean free path of 74 nm (55 nm) for an electron (a hole) in Au [18]. For a given Au-Si Schottky contact, $\Phi_B$ is fixed (i.e. $\Phi_B = 0.34$ and 0.8 eV for p-Si and n-Si interfaces, respectively [1]) and $N$ is determined by the photon energy $E_0$ and the Au film thickness $h_{Au}$. If $E_0$ (photon wavelength $\lambda$) becomes larger (smaller) and $h_{Au}$ becomes thinner, the hot carriers will undergo more round trips in the Au film (i.e., $N > 0$), leading to higher $P(E)$. This thus results in higher IQE according to Equation (3-2) and is clearly demonstrated in Figure 3(a) for p-Si based

Schottky contact. When $h_{Au}$ gets thick enough, $N$ will be reduced to zero and as a result $P(E)$ and IQE become independent of $h_{Au}$. In this case, $P(E) = P_{ii}$ is reduced to $P_i$ as expressed in Equation (3-1). The black dotted curve in Figure 3(a) shows where $N$ becomes zero. It shows that $h_{Au}|_{N=0}$ becomes thinner as $\lambda$ increases (or $E_0$ decreases). For n-Si based Schottky contact, IQE is about one order of magnitude smaller than that for the p-Si based Schottky contact as shown in Figure 3(b) and is independent of $h_{Au}$ in the whole wavelength range considered here. This is mainly due to the high $\Phi_B$ which prevents hot carriers excited by photons in this range to travel a complete round trip in the Au film even when its thickness is as small as 10 nm. As illustrated in Figure 2(b), $P(E)$ for Au-Si round interfaces follows the same expression, i.e., Equations (3-1) and (3-2), as that for flat interfaces. Actually, $P(E) = P_{ii}$ in Equation (3-2) is a unified formula including $P(E) = P_i$ in Equation (3-1) when $N = 0$.

For our detector shown in Figure 1(b), IQE can be calculated by employing the above $P(E)$ expressions. In detail, according to the assumption, hot carriers traveling only in the direction perpendicular to the Au-Si interface can have an opportunity for emission. Therefore, the Au slab (denoted by the red dashed rectangles, i.e., Region I) can be regarded to be a film thick enough horizontally and much larger than the dotted line shown in Figure 3(a) for the Au-p-Si interface. Thus, $P(E)$ in this region is mainly equal to $P_i$ in Equation (3-1). For the other part of the Au film, i.e., Region II, $h_{Au}$ is not as thick as that in Region I and becomes an important factor to influence $N$ in addition to the hot carrier energy $E_0$. Therefore, $P(E)$ is determined by either $P_i$ or $P_{ii}$ in Equation (3) according to the value of $N$.

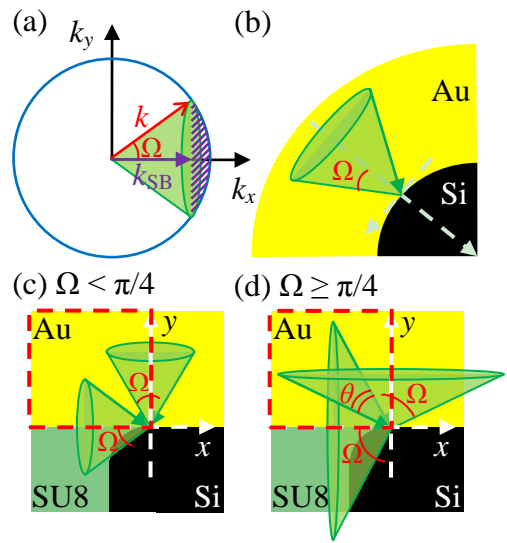

**Figure 2.** Emission cones with angle, $\Omega$: (a) a momentum-space illustration; (b) real-space

illustrations for round Au-Si interfaces and (c, d) an asymmetric corner denoted by the red dashed rectangles when (c) $\Omega \geq \pi/4$ and (d) $\Omega < \pi/4$.

For the structure with $h_{SU8}$ = 200 nm shown in Figure 1(c), the Au film also needs to be divided into two regions, i.e., Regions II and III, and $P(E)$ is calculated differently for each region. Region II, in direct contact with the top surface of the Si ridge waveguide, is the same as that shown in Figure 1(b), and $P(E)$ can be calculated there based on Equation (3). In Region III, where only a point on the corner is in direct contact with Si, as shown in Figure 1(c), the momentum of a hot carrier can be orthogonally decomposed into two components either along the horizontal direction or the vertical direction. This is illustrated as two emission cones along the $x$ and $y$ directions in Figures 2(c) and 2(d), respectively. The emission angle, $\Omega$, is determined by $k \cdot \cos\Omega = k_{SB}$ as shown in Figure 2(a). Similar to Regions I and II, the horizontal and vertical emission probabilities are mainly determined by $P_i$ and $P_{ii}$, respectively. Since only half of the emission cone lies in Region III, a hot carrier in each cone is only half as likely to undergo emission, that is, $P(E) = P_i/2$ when $N = 0$ and $P(E) = P_{ii}/2$ when $N > 0$. Since $P_{ii}$ of the vertical emission probability is not always equal to $P_i$ of the horizontal emission probability, the corner is actually asymmetrical. If these two cones are separated with a solid angle smaller than $\pi/4$, i.e., $\Omega \leq \pi/4$, as shown in Figure 2(c), the total IPE probability will be defined by the sum of the probabilities of each half cone divided by 2, that is,

$$P(E) = (P_i + P_{ii})/4 \qquad (4)$$

If $\Omega > \pi/4$, the two emission cones overlap as shown in Figure 2(d). The overlapping angle $\theta$ is equal to $(2\Omega - \pi/2)$. Then, the total IPE probability is the sum of the probabilities for the two half emission cones minus the overlapping probability and finally divided by 2. Since the vertical emission probability is not always equal to the horizontal one, it is not easy to analytically find the probability of a hot carrier in the overlapping region to emit in either direction. Here, we simply assume that all the hot carriers in the overlapping region have 50% probability from the vertical cone and 50% from the horizontal cone. Then, we get $P(E)$ as follows:

$$P(E) = (P_i + P_{ii} - P_i|_{\theta/2} - P_{ii}|_{\theta/2})/4 \qquad (5)$$

If $N = 0$ holds for the vertical emission, $P_{ii}$ is reduced to $P_i$ and $P(E)$ can be simplified as Equation (4). IQE in Region III is plotted in Figures 3(c) and 3(d) for p-Si and n-Si based Schottky diodes, respectively. For the n-Si based asymmetrical corner of Region III, $\Omega < \pi/4$ holds in the whole wavelength range due to the high Schottky barrier ($\Phi_B = 0.8$ eV) and it leads to just half of the values for the flat or round Au-Si contact shown in Figure 3(b). For p-Si based asymmetrical corner of Region III, IQE follows the same trend as that shown in Figure 3(a) for either the flat or round Au-Si contact. The turning point of $h_{Au}$ remains the same for all input wavelengths, as denoted by the black dotted line in Figure 3(c). Compared with the values shown in Figure 3(a), IQE at each point falls to about half, meaning that the overlap of the emission cones, shown in Figure 2(d), contributes little to the total IPE probability.

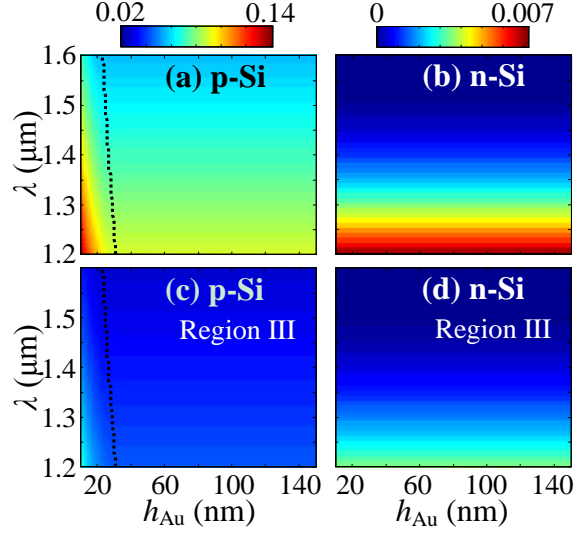

**Figure 3.** IQE as a function of the Au electrode thickness $h_{Au}$ and the incident photon wavelength $\lambda$ for (a) p-Si and (b) n-Si based Au-Si flat or round interfaces; (c) p-Si and (d) n-Si based asymmetrical corner cases in Region III. The black dotted curves shown in Figures (a) and (c) indicated where IQE becomes independent of $h_{Au}$.

Like IQE, the absorption, $A$, also differs for different regions and it can be calculated based on the finite-difference time domain method using the commercial software Lumerical FDTD Solutions. The fundamental TE or TM modes of the dielectric Si ridge waveguide are set as the input. Three-dimensional monitors are set to record the electric field intensity distribution ($|E|^2$) of different parts of the Au film. The absorption can be obtained as follows:

$$A = \frac{\iiint \frac{\pi c}{\lambda} \cdot \mathrm{Im}(\varepsilon) |E|^2 \, dxdydz}{\text{source power}}, \quad (6)$$

where $\mathrm{Im}(\varepsilon)$ is the imaginary part of the dielectric constant of Au, and $c$ is the speed of light in vacuum. To calculate the total responsivities of photodetectors shown in

Figures 1(b) and 1(c), we extended Equation (1) into the sum of responsivities in compositional regions as expressed in Equations (7) and (8) below:

$$R = \frac{q}{E_0} \cdot \left( A_\text{I} \cdot IQE_\text{I} + A_\text{II} \cdot IQE_\text{II} \right) \quad (7)$$

$$R = \frac{q}{E_0} \cdot \left( A_\text{II} \cdot IQE_\text{II} + A_\text{III} \cdot IQE_\text{III} \right) \quad (8)$$

where, $A_x \cdot IQE_x$ is the external quantum efficiency in Region x. Equations (7) and (8) mean that the photocurrent collected by the electrode is contributed by all the hot carriers emitted from all the compositional regions: Regions I and II in Figure 1(b) and Regions II and III in Figure 1(c).

For a PD, dark current is also a very important parameter, which determines the sensitivity or the smallest detectable power of a PD. For a Schottky diode, the dark current can be calculated using Equation (9) given below [16]:

$$I_{dark} = C_{area} A^{**} T^2 e^{-\left( q\Phi_B / k_B T \right)} \quad (9)$$

where $C_\text{area}$ is the Schottky contact area, $A^{**}$ is the effective Richardson constant, equal to 112 and 32 $\text{Acm}^{-2}\text{K}^{-2}$ for electrons and holes, respectively [21], $k_\text{B} = 1.3806505 \times 10^{-23}$ J/K is the Boltzmann constant and $T = 300$ K at room temperature.

In the following, only the p-Si based Schottky diode (instead of the n-Si based one) is considered because of its much higher IQE shown in Figure 3, and is therefore promising to produce a much larger photocurrent with the same diode structure. Since $h_\text{SU8}$ plays a very important role in tuning the Au sidewall coverage and further tuning the absorption polarization dependence, we first investigated the responsivity $R$ of the photodetector as a function of $h_\text{SU8}$. The curves representing $h_\text{SU8} = 50, 60, 70, 80, 90, 100$ and $200$ nm are plotted in Figures 4(a) and 4(b) for TE and TM polarizations, respectively. From these two figures, it is seen that with the Au film covering part of the sidewalls of the Si ridge waveguide (i.e., $h_\text{SU8} < 200$ nm schematically shown in Figure 1(b)), $R$ becomes much higher than that for the case of $h_\text{SU8} = 200$ nm (i.e., no sidewall coverage of Au, as schematically shown in Figure 1(c)). For TE polarization, $R$ decreases linearly as the wavelength increases in the whole wavelength range of 1.2-1.6 μm when $h_\text{SU8} \leq 60$ nm. As $h_\text{SU8}$ increases to 100 nm, a responsivity peak starts to appear and then red-shifts gradually as shown in Figure 4(a). For TM polarization, the $R$ spectrum falls gradually when $h_\text{SU8}$ increases to 100 nm as shown in Figure 4(b). Interference ripples are observed clearly in the short wavelength range. To quantitatively characterize the polarization sensitivity, we define polarization dependent deviations (PDD) of responsivity between TE and TM polarizations as follows,

$$PDD = \frac{|R_{TE} - R_{TM}|}{(R_{TE} + R_{TM})/2} \times 100\% \qquad (10)$$

Based on Figures 4(a) and 4(b), PDD is calculated according to Equation (10) and plotted in Figure 4(c) for different values of $h_{SU8}$. It is shown that the PDD spectrum for the case of $h_{SU8}$ = 60 nm is the lowest (below 1%) in the wavelength range of 1.2-1.5 μm, beyond which the value of PDD rises quickly with the increasing $\lambda$. As $h_{SU8}$ increases, higher polarization sensitivities appear, leading to the rising PDD spectrum. PDD approaches a maximum in the short wavelength range when $h_{SU8}$ = 200 nm (i.e., the comparative case when there is no coverage of Au on the sidewall of the Si ridge waveguide). This indicates that a certain degree of coverage of the Au film on the sidewall of the Si ridge waveguide is also beneficial to minimize the polarization sensitivity of the responsivity.

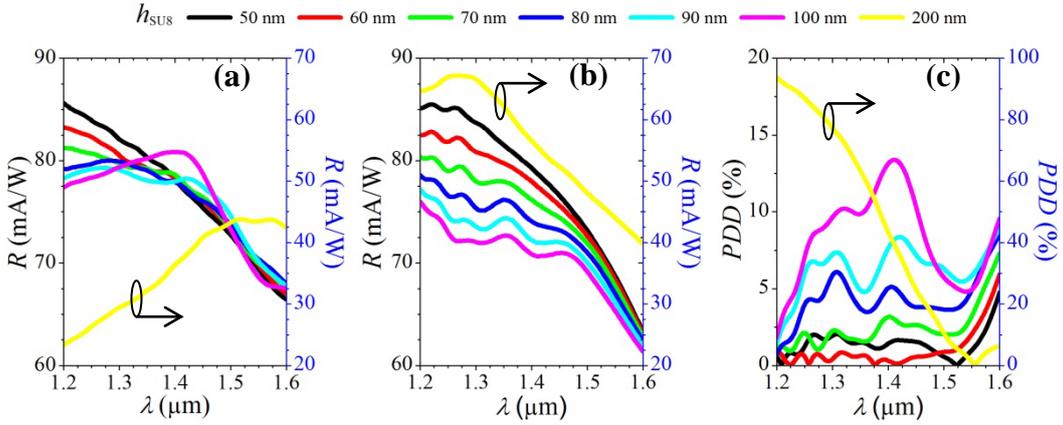

**Figure 4.** Responsivity varies with the SU8 film thickness, $h_{SU8}$ for (a) TE and (b) TM polarizations. (c) Polarization dependent deviations of responsivity, *PDD*. $h_{SU8}$ = 50, 60, 70, 80, 90, 100, 200 nm are considered and the Au film thickness is fixed as $h_{Au}$ = 20 nm. In all the three figures, the comparative curves representing the case of $h_{SU8}$ = 200 nm (i.e., the comparative case of Fig. 1(c)) are all calibrated by the right vertical axes to give clear comparison.

In order to have a more detailed understanding of the polarization tuning, the total and partial absorption, reflection, and transmission spectra of two typical cases of $h_{SU8}$ = 60 and 200 nm are investigated, as demonstrated in Figure 5. From the total absorption spectra shown in Figure 5(a), it is seen that for the comparative detector of $h_{SU8}$ = 200 nm, the absorption for TE polarization is lower than that for TM polarization. Both of these spectra are much lower than that of our detector with $h_{SU8}$

= 60 nm with absorption of greater than 85% and apparently a very weak polarization deviation in the whole wavelength range. This indicates that the sidewall coverage of Au greatly enhances the total absorption, especially for the TE polarization, leaving little light passing through the Au Schottky contact region as shown in Figure 5(b). It is noted that it does not cause much reflection, as shown in Figure 5(a). To investigate the regional optical responses, partial absorption in Regions I-III for TE and TM polarizations are calculated and plotted in Figures 5(c) and 5(d), which show how much absorption and consequently the photocurrent each part of the Au film contributes to. For TE polarization, in comparison with the absorption spectra in Region II, as shown in Figure 5(d), the absorption in Region I, i.e., $A_I$, is much higher for our photodetector of $h_{SU8}$ = 60 nm and is also much higher than that in Region III, i.e., $A_{III}$ for the detector of $h_{SU8}$ = 200 nm as shown in Figure 5(c). This indicates that more TE-polarized light is confined in the Au slab being absorbed and the Au slab absorption is significantly enhanced with the Au sidewall. By checking the electric field intensity distribution (not shown here), we found that the electric field is mainly confined around the Au-Si contacting interfaces. For TM polarization, the absorption spectra do not deviate much from each other in Region I (or III) and Region II. The top and side parts of the Au film absorb more light than the Au slab at short wavelengths below about 1.4 μm for the TM polarization. This is because the optical field is mainly confined vertically between the top Au film and the bottom of the Si ridge waveguide, and light is mainly absorbed by the top part of the Au film. In the long wavelength range, the optical field extends horizontally into the Au slab leading to increasing $A_I$ and $A_{III}$ (i.e., light absorption in Regions I and III) but decreasing $A_{II}$ (i.e., absorption in Region II). Their complementary behaviors make the total absorption spectrum for TM polarization flat as shown in Figure 5(a). At $h_{Au}$ = 20 nm, IQE in Region II and Region I (or III) are different as shown in Figures 3(a) and 3(c), respectively, but the responsivity spectra of both TE and TM polarizations do not deviate much from the trend of the absorption shown in Figure 5(a) but linearly decrease with IQE.

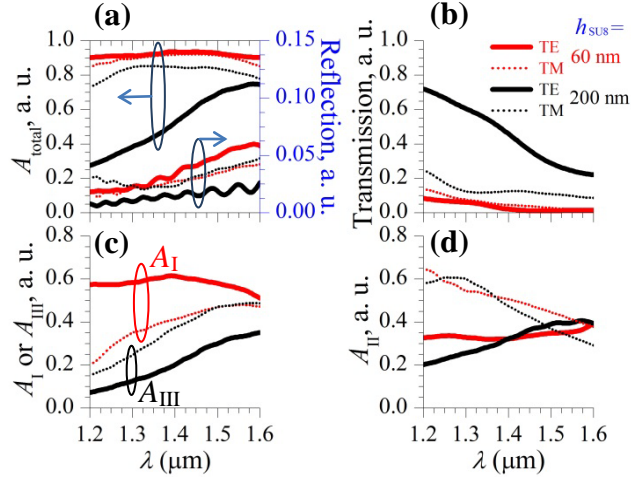

**Figure 5.** Comparisons of Schottky diodes of $h_{SU8}$ = 60 (red) and 200 nm (black) for TE (solid lines) and TM (dotted lines) polarizations on (a) total absorption and reflection (which is calibrated by the right vertical axis); (b) Transmission; (c) partial absorption, $A_I$ for $h_{SU8}$ = 60 nm and $A_{III}$ for $h_{SU8}$ = 200 nm; and (f) partial absorption, $A_{II}$. The Au film thickness is $h_{Au}$ = 20 nm.

Another important parameter is the Au film thickness, which influences not only the IQE (shown in Figure 3) but also the light absorption. Here, we fix $h_{SU8}$ = 60 nm (with which *PDD* can be minimized the most significantly when $h_{Au}$ = 20 nm as shown in Figure 4) and investigate the effect of $h_{Au}$ on the responsivity of our detector. Figure 6(a) shows the total absorption spectra of photodetectors with $h_{Au}$ = 10, 20, 30, 50, 70, and 90 nm. When $h_{Au} \geq 50$ nm, the absorption spectrum tends not to change much with increasing $h_{Au}$ for both TE and TM polarizations. This is because of the tightly confined optical field at the Au-Si interface that can be sufficiently absorbed with a 50-nm thick Au film. Also due to the unchanged IQE shown in Figure 3, *R* in Figure 6(b) and *PDD* in Figure 6(c), consequently, do not show any change with $h_{Au}$. However, *PDD* is very large in the whole wavelength range. As $h_{Au}$ decreases, *PDD* decreases first and then quickly increases as shown in Figure 6(c). At $h_{Au}$ = 20 nm, *PDD* reaches the minimal value in most wavelengths from 1.2 to 1.5 μm. As shown in Figure 6(a), for TE polarization, an absorption peak is observed for each $h_{Au}$ and it red-shifts as $h_{Au}$ decreases. With the linearly decreasing IQE as *λ* increases in each part of the Au film, as shown in Figure 3, the peak in responsivity becomes weakened as shown in Figure 6(b). For TM polarization, the absorption spectrum moves up until $h_{Au}$ decreases to 20 nm as shown in Figure 6(a). Consequently, the responsivity spectrum in Figure 6(b) also moves up and moves extremely quickly when $h_{Au}$ becomes below 30 nm, where IQE increases much more quickly. At $h_{Au}$ = 10 nm, with the even greater IQE, the responsivity spectrum goes up even higher even though the absorption spectrum is below that for the case of $h_{Au}$ = 20 nm. Then, the IQE dominates the responsivity spectrum.

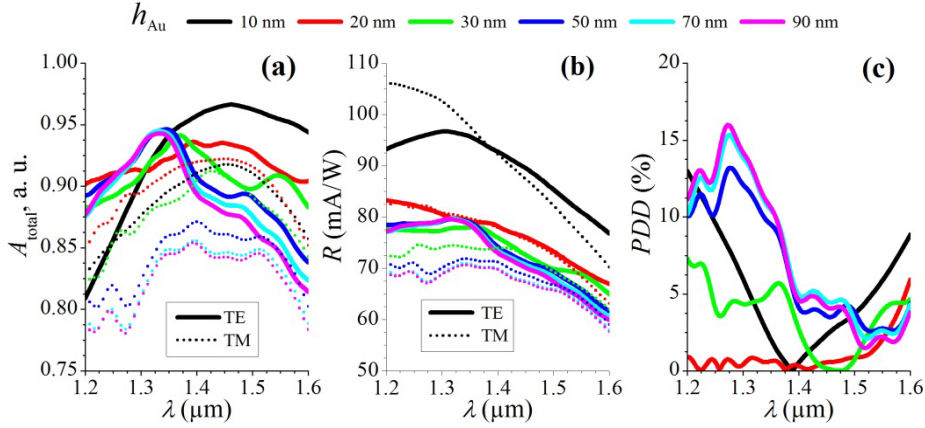

**Figure 6.** (a) Total absorption, $A_{total}$, (b) responsivity, $R$, for both TE (solid lines) and TM (dotted lines) polarizations, and (c) polarization dependent deviation $PDD$ of responsivity, for photodetectors with $h_{Au}$ = 10, 20, 30, 50, 70, and 90 nm. The SU8 film thickness is $h_{SU8}$ = 60 nm.

In summary, we have proposed a new design for a silicon Schottky photodiode shown in Figure 1(a). Based on the extensive physical model studied here, we found that the polarization insensitivity can be greatly minimized by our photodetector with $h_{SU8}$ = 60 nm and $h_{Au}$ = 20 nm in the optical communication wavelength range of 1.2 – 1.5 μm. It has much higher responsivity than the one with the same Schottky contact length shown in Figure 1(c). The polarization sensitivity of responsivity has also been greatly minimized. All of these excellent properties are attributed to the highly-absorptive plasmonic waveguide with sidewall coverage of Au. Over 85% of the input light can be absorbed by an only 4-μm long Au film, which is much smaller than those reported in Refs. [9-11]. Because of this, its dark current calculated by Equation (10) is also very low, i.e., $I_{dark}$ = 136 nA, much smaller than those reported in Refs. [9, 10]. Therefore, the new design can be effectively used for applications in high-density silicon photonic integration.

## ACKNOWLEDGMENT
This work was supported by the Zhejiang Provincial Key Project (No. 2011C11024) and National Natural Science Foundation (Nos. 91233208, 91233119, and 61307078).